








   \def\bbbc{{\mathchoice {\setbox0=\hbox{$\displaystyle\rm C$}\hbox{\hbox
   to0pt{\kern0.4\wd0\vrule height0.9\ht0\hss}\box0}}
   {\setbox0=\hbox{$\textstyle\rm C$}\hbox{\hbox
   to0pt{\kern0.4\wd0\vrule height0.9\ht0\hss}\box0}}
   {\setbox0=\hbox{$\scriptstyle\rm C$}\hbox{\hbox
   to0pt{\kern0.4\wd0\vrule height0.9\ht0\hss}\box0}}
   {\setbox0=\hbox{$\scriptscriptstyle\rm C$}\hbox{\hbox
   to0pt{\kern0.4\wd0\vrule height0.9\ht0\hss}\box0}}}}

   \def\bbbg{{\mathchoice {\setbox0=\hbox{$\displaystyle\rm G$}\hbox{\hbox
   to0pt{\kern0.4\wd0\vrule height0.9\ht0\hss}\box0}}
   {\setbox0=\hbox{$\textstyle\rm G$}\hbox{\hbox
   to0pt{\kern0.4\wd0\vrule height0.9\ht0\hss}\box0}}
   {\setbox0=\hbox{$\scriptstyle\rm G$}\hbox{\hbox
   to0pt{\kern0.4\wd0\vrule height0.9\ht0\hss}\box0}}
   {\setbox0=\hbox{$\scriptscriptstyle\rm G$}\hbox{\hbox
   to0pt{\kern0.4\wd0\vrule height0.9\ht0\hss}\box0}}}}

   \def\bbbo{{\mathchoice {\setbox0=\hbox{$\displaystyle\rm O$}\hbox{\raise
   0.15\ht0\hbox to0pt{\kern0.4\wd0\vrule height0.8\ht0\hss}\box0}}
   {\setbox0=\hbox{$\textstyle\rm O$}\hbox{\raise
   0.15\ht0\hbox to0pt{\kern0.4\wd0\vrule height0.8\ht0\hss}\box0}}
   {\setbox0=\hbox{$\scriptstyle\rm O$}\hbox{\raise
   0.15\ht0\hbox to0pt{\kern0.4\wd0\vrule height0.7\ht0\hss}\box0}}
   {\setbox0=\hbox{$\scriptscriptstyle\rm O$}\hbox{\raise
   0.15\ht0\hbox to0pt{\kern0.4\wd0\vrule height0.7\ht0\hss}\box0}}}}

   \def\bbbq{{\mathchoice {\setbox0=\hbox{$\displaystyle\rm Q$}\hbox{\raise
   0.15\ht0\hbox to0pt{\kern0.4\wd0\vrule height0.8\ht0\hss}\box0}}
   {\setbox0=\hbox{$\textstyle\rm Q$}\hbox{\raise
   0.15\ht0\hbox to0pt{\kern0.4\wd0\vrule height0.8\ht0\hss}\box0}}
   {\setbox0=\hbox{$\scriptstyle\rm Q$}\hbox{\raise
   0.15\ht0\hbox to0pt{\kern0.4\wd0\vrule height0.7\ht0\hss}\box0}}
   {\setbox0=\hbox{$\scriptscriptstyle\rm Q$}\hbox{\raise
   0.15\ht0\hbox to0pt{\kern0.4\wd0\vrule height0.7\ht0\hss}\box0}}}}

   \def\bbbt{{\mathchoice {\setbox0=\hbox{$\displaystyle\rm
   T$}\hbox{\hbox to0pt{\kern0.3\wd0\vrule height0.9\ht0\hss}\box0}}
   {\setbox0=\hbox{$\textstyle\rm T$}\hbox{\hbox
   to0pt{\kern0.3\wd0\vrule height0.9\ht0\hss}\box0}}
   {\setbox0=\hbox{$\scriptstyle\rm T$}\hbox{\hbox
   to0pt{\kern0.3\wd0\vrule height0.9\ht0\hss}\box0}}
   {\setbox0=\hbox{$\scriptscriptstyle\rm T$}\hbox{\hbox
   to0pt{\kern0.3\wd0\vrule height0.9\ht0\hss}\box0}}}}

   \def\bbbs{{\mathchoice
   {\setbox0=\hbox{$\displaystyle     \rm S$}\hbox{\raise0.5\ht0\hbox
   to0pt{\kern0.35\wd0\vrule height0.45\ht0\hss}\hbox
   to0pt{\kern0.55\wd0\vrule height0.5\ht0\hss}\box0}}
   {\setbox0=\hbox{$\textstyle        \rm S$}\hbox{\raise0.5\ht0\hbox
   to0pt{\kern0.35\wd0\vrule height0.45\ht0\hss}\hbox
   to0pt{\kern0.55\wd0\vrule height0.5\ht0\hss}\box0}}
   {\setbox0=\hbox{$\scriptstyle      \rm S$}\hbox{\raise0.5\ht0\hbox
   to0pt{\kern0.35\wd0\vrule height0.45\ht0\hss}\raise0.05\ht0\hbox
   to0pt{\kern0.5\wd0\vrule height0.45\ht0\hss}\box0}}
   {\setbox0=\hbox{$\scriptscriptstyle\rm S$}\hbox{\raise0.5\ht0\hbox
   to0pt{\kern0.4\wd0\vrule height0.45\ht0\hss}\raise0.05\ht0\hbox
   to0pt{\kern0.55\wd0\vrule height0.45\ht0\hss}\box0}}}}

   \def\bbbu{{\mathchoice {\setbox0=\hbox{$\displaystyle\rm U$}\hbox{\hbox
   to0pt{\kern0.4\wd0\vrule height0.9\ht0\hss}\box0}}
   {\setbox0=\hbox{$\textstyle\rm U$}\hbox{\hbox
   to0pt{\kern0.4\wd0\vrule height0.9\ht0\hss}\box0}}
   {\setbox0=\hbox{$\scriptstyle\rm U$}\hbox{\hbox
   to0pt{\kern0.4\wd0\vrule height0.9\ht0\hss}\box0}}
   {\setbox0=\hbox{$\scriptscriptstyle\rm U$}\hbox{\hbox
   to0pt{\kern0.4\wd0\vrule height0.9\ht0\hss}\box0}}}}

   \def\bbbz{{\mathchoice {\hbox{$\textstyle Z\kern-0.4em Z$}}
   {\hbox{$\textstyle Z\kern-0.4em Z$}}
   {\hbox{$\scriptstyle Z\kern-0.3em Z$}}
   {\hbox{$\scriptscriptstyle Z\kern-0.2em Z$}}}}



      \font \ninebf                 = cmbx9
      \font \ninei                  = cmmi9
      \font \nineit                 = cmti9
      \font \ninerm                 = cmr9
      \font \ninesans               = cmss10 at 9pt
      \font \ninesl                 = cmsl9
      \font \ninesy                 = cmsy9
      \font \ninett                 = cmtt9
      \font \fivesans               = cmss10 at 5pt
                                                \font \sevensans              = cmss10 at 7pt
      \font \sixbf                  = cmbx6
      \font \sixi                   = cmmi6
      \font \sixrm                  = cmr6
                                                \font \sixsans                = cmss10 at 6pt
      \font \sixsy                  = cmsy6
      \font \tams                   = cmmib10
      \font \tamss                  = cmmib10 scaled 700
                                                \font \tensans                = cmss10
    
      \skewchar\ninei='177 \skewchar\sixi='177
      \skewchar\ninesy='60 \skewchar\sixsy='60
      \hyphenchar\ninett=-1
      \def\newline{\hfil\break}%
      \catcode`@=11
      \def\folio{\ifnum\pageno<\z@
      \uppercase\expandafter{\romannumeral-\pageno}%
      \else\number\pageno \fi}
      \catcode`@=12 

      \newfam\sansfam
      \textfont\sansfam=\tensans\scriptfont\sansfam=\sevensans
      \scriptscriptfont\sansfam=\fivesans
      \def\sans{\fam\sansfam\tensans}
\def\tens#1{\relax\ifmmode
\mathchoice{\hbox{$\displaystyle\sans#1$}}{\hbox{$\textstyle\sans#1$}}
{\hbox{$\scriptstyle\sans#1$}}{\hbox{$\scriptscriptstyle\sans#1$}}\else
$\sans#1$\fi}



      \def\petit{\def\rm{\fam0\ninerm}%
      \textfont0=\ninerm \scriptfont0=\sixrm \scriptscriptfont0=\fiverm
       \textfont1=\ninei \scriptfont1=\sixi \scriptscriptfont1=\fivei
       \textfont2=\ninesy \scriptfont2=\sixsy \scriptscriptfont2=\fivesy
       \def\it{\fam\itfam\nineit}%
       \textfont\itfam=\nineit
       \def\sl{\fam\slfam\ninesl}%
       \textfont\slfam=\ninesl
       \def\bf{\fam\bffam\ninebf}%
       \textfont\bffam=\ninebf \scriptfont\bffam=\sixbf
       \scriptscriptfont\bffam=\fivebf
       \def\sans{\fam\sansfam\ninesans}%
       \textfont\sansfam=\ninesans \scriptfont\sansfam=\sixsans
       \scriptscriptfont\sansfam=\fivesans
       \def\tt{\fam\ttfam\ninett}%
       \textfont\ttfam=\ninett
       \normalbaselineskip=11pt
       \setbox\strutbox=\hbox{\vrule height7pt depth2pt width0pt}%
       \normalbaselines\rm


      \def\bvec##1{{\textfont1=\tbms\scriptfont1=\tbmss
      \textfont0=\ninebf\scriptfont0=\sixbf
      \mathchoice{\hbox{$\displaystyle##1$}}{\hbox{$\textstyle##1$}}
      {\hbox{$\scriptstyle##1$}}{\hbox{$\scriptscriptstyle##1$}}}}}


\font\teneufm=eufm10
\font\seveneufm=eufm7
\font\fiveeufm=eufm5
\newfam\eufmfam
\textfont\eufmfam=\teneufm
\scriptfont\eufmfam=\seveneufm
\scriptscriptfont\eufmfam=\fiveeufm



                                        \mathchardef\Gammav="0100
     \mathchardef\Deltav="0101
     \mathchardef\Thetav="0102
     \mathchardef\Lambdav="0103
     \mathchardef\Xiv="0104
     \mathchardef\Piv="0105
     \mathchardef\Sigmav="0106
     \mathchardef\Upsilonv="0107
     \mathchardef\Phiv="0108
     \mathchardef\Psiv="0109
     \mathchardef\Omegav="010A


                                        \mathchardef\Gammau="0000
     \mathchardef\Deltau="0001
     \mathchardef\Thetau="0002
     \mathchardef\Lambdau="0003
     \mathchardef\Xiu="0004
     \mathchardef\Piu="0005
     \mathchardef\Sigmau="0006
     \mathchardef\Upsilonu="0007
     \mathchardef\Phiu="0008
     \mathchardef\Psiu="0009
     \mathchardef\Omegau="000A


\font\grbfivefm=cmbx5
\font\grbsevenfm=cmbx7
\font\grbtenfm=cmbx10 
\newfam\grbfam
\textfont\grbfam=\grbtenfm
\scriptfont\grbfam=\grbsevenfm
\scriptscriptfont\grbfam=\grbfivefm

\font\calbfivefm=cmbsy10 at 5pt
\font\calbsevenfm=cmbsy10 at 7pt
\font\calbtenfm=cmbsy10 
\newfam\calbfam
\textfont\calbfam=\calbtenfm
\scriptfont\calbfam=\calbsevenfm
\scriptscriptfont\calbfam=\calbfivefm



      \def\bvec#1{{\textfont1=\tams\scriptfont1=\tamss
      \textfont0=\tenbf\scriptfont0=\sevenbf
      \mathchoice{\hbox{$\displaystyle#1$}}{\hbox{$\textstyle#1$}}
      {\hbox{$\scriptstyle#1$}}{\hbox{$\scriptscriptstyle#1$}}}}


\def\pmbf#1{\leavevmode\setbox0=\hbox{#1}%
\kern-.025em\copy0\kern-\wd0
\kern.05em\copy0\kern-\wd0
\kern-.025em\copy0\kern-\wd0
\kern-.04em\copy0\kern-\wd0
\kern.08em\box0 }



                                                \def\monthname{%
                        \ifcase\month
      \or Jan\or Feb\or Mar\or Apr\or May\or Jun%
      \or Jul\or Aug\or Sep\or Oct\or Nov\or Dec%
                        \fi
                                                        }%
                                        \def\timestring{\begingroup
                \count0 = \time
                \divide\count0 by 60
                \count2 = \count0   
                \count4 = \time
                \multiply\count0 by 60
                \advance\count4 by -\count0   
                \ifnum\count4<10
     \toks1 = {0}%
                \else
     \toks1 = {}%
                \fi
                \ifnum\count2<12
      \toks0 = {a.m.}%
                \else
      \toks0 = {p.m.}%
      \advance\count2 by -12
                \fi
                \ifnum\count2=0
      \count2 = 12
                \fi
                \number\count2:\the\toks1 \number\count4 \thinspace \the\toks0
                                        \endgroup}%

                                \newskip\abovelistskip      \abovelistskip = .5\baselineskip 
                                \newskip\interitemskip      \interitemskip = 0pt
                                \newskip\belowlistskip      \belowlistskip = .5\baselineskip
                                \newdimen\listleftindent    \listleftindent = 0pt
                                \newdimen\listrightindent   \listrightindent = 0pt

                                %

                                                

                                                \def\mes{%
                        \ifcase\month
      \or Enero\or Febrero\or Marzo\or Abril\or Mayo\or Junio%
      \or Julio\or Agosto\or Septiembre\or Octubre\or Noviembre\or Diciembre%
                        \fi
                                                        }%
                                        \def\timestring{\begingroup
                \count0 = \time
                \divide\count0 by 60
                \count2 = \count0   
                \count4 = \time
                \multiply\count0 by 60
                \advance\count4 by -\count0   
                \ifnum\count4<10
     \toks1 = {0}%
                \else
     \toks1 = {}%
                \fi
                \ifnum\count2<12
      \toks0 = {a.m.}%
                \else
      \toks0 = {p.m.}%
      \advance\count2 by -12
                \fi
                \ifnum\count2=0
      \count2 = 12
                \fi
                \number\count2:\the\toks1 \number\count4 \thinspace \the\toks0
                                        \endgroup}%


                                %
                                \newskip\abovelistskip      \abovelistskip = .5\baselineskip 
                                \newskip\interitemskip      \interitemskip = 0pt
                                \newskip\belowlistskip      \belowlistskip = .5\baselineskip
                                \newdimen\listleftindent    \listleftindent = 0pt
                                \newdimen\listrightindent   \listrightindent = 0pt

                                %
      \def\petit{\def\rm{\fam0\ninerm}%
      \textfont0=\ninerm \scriptfont0=\sixrm \scriptscriptfont0=\fiverm
       \textfont1=\ninei \scriptfont1=\sixi \scriptscriptfont1=\fivei
       \textfont2=\ninesy \scriptfont2=\sixsy \scriptscriptfont2=\fivesy
       \def\it{\fam\itfam\nineit}%
       \textfont\itfam=\nineit
       \def\sl{\fam\slfam\ninesl}%
       \textfont\slfam=\ninesl
       \def\bf{\fam\bffam\ninebf}%
       \textfont\bffam=\ninebf \scriptfont\bffam=\sixbf
       \scriptscriptfont\bffam=\fivebf
       \def\sans{\fam\sansfam\ninesans}%
       \textfont\sansfam=\ninesans \scriptfont\sansfam=\sixsans
       \scriptscriptfont\sansfam=\fivesans
       \def\tt{\fam\ttfam\ninett}%
       \textfont\ttfam=\ninett
       \normalbaselineskip=11pt
       \setbox\strutbox=\hbox{\vrule height7pt depth2pt width0pt}%
       \normalbaselines\rm
      \def\vec##1{{\textfont1=\tbms\scriptfont1=\tbmss
      \textfont0=\ninebf\scriptfont0=\sixbf
      \mathchoice{\hbox{$\displaystyle##1$}}{\hbox{$\textstyle##1$}}
      {\hbox{$\scriptstyle##1$}}{\hbox{$\scriptscriptstyle##1$}}}}}


      \def\footnoterule{\kern-3pt\hrule width 2cm\kern2.6pt}
      \newdimen\oldparindent\oldparindent=1.5em
      \parindent=1.5em
 
      \newcount\footcount \footcount=0
      \def\advftncnt{\advance\footcount by1\global\footcount=\footcount}
      \def\fnote#1{\advftncnt$^{\the\footcount}$\begingroup\petit
      \parfillskip=0pt plus 1fil
      \def\textindent##1{\hangindent0.5\oldparindent\noindent\hbox
      to0.5\oldparindent{##1\hss}\ignorespaces}%
      \vfootnote{$^{\the\footcount}$}{#1\nullbox{0mm}{2mm}{0mm}\vskip-9.69pt}\endgroup}


      \def\item#1{\par\noindent
      \hangindent6.5 mm\hangafter=0
      \llap{#1\enspace}\ignorespaces}
      
      \def\leaderfill{\kern0.5em\leaders\hbox to 0.5em{\hss.\hss}\hfill\kern
      0.5em}
                                                \def\hb{\hfill\break}

    \def\centerrule#1{\centerline{\kern#1\hrulefill\kern#1}}


      \def\boxit#1{\vbox{\hrule\hbox{\vrule\kern3pt
                                                \vbox{\kern3pt#1\kern3pt}\kern3pt\vrule}\hrule}}

      \def\tightboxit#1{\vbox{\hrule\hbox{\vrule
                                                \vbox{#1}\vrule}\hrule}}

      \def\looseboxit#1{\vbox{\hrule\hbox{\vrule\kern5pt
                                                \vbox{\kern5pt#1\kern5pt}\kern5pt\vrule}\hrule}}

      \def\youboxit#1#2{\vbox{\hrule\hbox{\vrule\kern#2
                                                \vbox{\kern#2#1\kern#2}\kern#2\vrule}\hrule}}



                        \def\whitetile#1#2#3{\setbox0=\null
                        \ht0=#1 \dp0=#2\wd0=#3 \setbox1=\hbox{\tightboxit{\box0}}\lower#2\box1}

                        \def\nullbox#1#2#3{\setbox0=\null
                        \ht0=#1 \dp0=#2\wd0=#3\box0}


   \def\permil{\leavevmode\hbox{\raise1ex%
   \hbox{$\scriptscriptstyle0$}\kern-0.2em%
   \raise0.4ex\hbox{\rm\char"2F}\kern-0.2em\hbox{$\scriptscriptstyle00$}}}




\def\equ{\leavevmode Eq.}

\def\equs{\leavevmode Eqs.}

\def\equn#1{\ifmmode \eqno{\rm #1}\else \equ~#1\fi}

\def\ecu{\leavevmode Ec.}

\def\ecun#1{\ifmmode \eqno{\rm #1}\else \ecu~#1\fi}


\def\gev{\ifmmode \mathop{\rm GeV}\nolimits\else {\rm GeV}\fi}
\def\mev{\ifmmode \mathop{\rm MeV}\nolimits\else {\rm MeV}\fi}
\def\kev{\ifmmode \mathop{\rm keV}\nolimits\else {\rm keV}\fi}
\def\ev{\ifmmode \mathop{\rm eV}\nolimits\else {\rm eV}\fi}

\def\erg{\ifmmode \mathop{\rm erg}\nolimits\else {\rm erg}\fi}
\def\ryd{\ifmmode \mathop{\rm Ry}\nolimits\else {\rm Ry}\fi}
\def\angst{\ifmmode\mathop{\rm\AA}\nolimits\else {\rm \AA}\fi}

\def\degreec{\ifmmode\mathop{^\circ \rm C}\nolimits\else{$^\circ{\rm C}\;$}\fi}
\def\degreek{\ifmmode\mathop{^\circ \rm K}\nolimits\else{$^\circ{\rm K}\;$}\fi}
\def\degreef{\ifmmode\mathop{^\circ \rm F}\nolimits\else{$^\circ{\rm F}\;$}\fi}

\def\chidof{\ifmmode\mathop\chi^2/{\rm d.o.f.}\else $\chi^2/{\rm d.o.f.}\;$\fi}

\def\msbar{\hbox{$\overline{\hbox{MS}}$}}

\def\cmass{\ifmmode\mathop{\rm c.m.}\nolimits\else {\sl c.m.}\fi}
\def\lab{\ifmmode{\rm lab}\else {\sl lab.}\fi}

\def\degrees{\ifmmode{^\circ\,}\else $^\circ$\fi}
\def\feet{\ifmmode{\hbox{'}\,}\else '\fi}
\def\inches{\ifmmode{\hbox{"}\,}\else "\fi}



\def\TeX{T\kern-.1667em\lower.5ex\hbox{E}\kern-.125emX\null}%
\def\LaTeX{L\kern -.26em \raise .6ex \hbox{\sevenrm A}\kern -.15em \TeX}%
\def\AMSTeX{$\cal A\kern -.1667em
   \lower .5ex\hbox{$\cal M$}%
   \kern -.125em S$-\TeX
}%
\def\BibTeX{{\rm B\kern-.05em{\sevenrm I\kern-.025em B}\kern-.08em
    T\kern-.1667em\lower.7ex\hbox{E}\kern-.125emX}}%
\def\physmatex{P\kern-.14em\lower.5ex\hbox{\sevenrm H}ys
\kern -.35em \raise .6ex \hbox{{\sevenrm M}a}\kern -.15em
 T\kern-.1667em\lower.5ex\hbox{E}\kern-.125emX\null}%

\def\ref#1{$^{[#1]}$\relax}











\def\dddoverdots{.\kern-.8pt .\kern-.8pt .}


\def\underdot#1{\mathord{\vtop to0pt{\ialign{##\crcr
$\hfil\displaystyle{#1}\hfil$\crcr\noalign{\kern1.5pt\nointerlineskip}
$\hfil\dot{}\kern1.5pt\hfil$\crcr}\vss}}}

\def\underddot#1{\mathord{\vtop to0pt{\ialign{##\crcr
$\hfil\displaystyle{#1}\hfil$\crcr\noalign{\kern1.5pt\nointerlineskip}
$\hfil{.\kern-0.7pt.}{}\kern1.5pt\hfil$\crcr}\vss}}}

\def\underdddot#1{\mathord{\vtop to0pt{\ialign{##\crcr
$\hfil\displaystyle{#1}\hfil$\crcr\noalign{\kern1.5pt\nointerlineskip}
$\hfil{.\kern-0.7pt .\kern-0.7pt .}{}\kern1.5pt\hfil$\crcr}\vss}}}


\def\undertilde#1{\mathord{\vtop to0pt{\ialign{##\crcr
$\hfil\displaystyle{#1}\hfil$\crcr\noalign{\kern1.5pt\nointerlineskip}
$\hfil\tilde{}\kern1.5pt\hfil$\crcr}\vss}}}


\def\slash#1{#1\!\!\!/\,}



\def\ddal{\mathop{\vrule height 7pt depth0.2pt
\hbox{\vrule height 0.5pt depth0.2pt width 6.2pt}\vrule height 7pt depth0.2pt width0.8pt
\kern-7.4pt\hbox{\vrule height 7pt depth-6.7pt width 7.pt}}}
\def\sdal{\mathop{\kern0.1pt\vrule height 4.9pt depth0.15pt
\hbox{\vrule height 0.3pt depth0.15pt width 4.6pt}\vrule height 4.9pt depth0.15pt width0.7pt
\kern-5.7pt\hbox{\vrule height 4.9pt depth-4.7pt width 5.3pt}}}
\def\ssdal{\mathop{\kern0.1pt\vrule height 3.8pt depth0.1pt width0.2pt
\hbox{\vrule height 0.3pt depth0.1pt width 3.6pt}\vrule height 3.8pt depth0.1pt width0.5pt
\kern-4.4pt\hbox{\vrule height 4pt depth-3.9pt width 4.2pt}}}


\def\dlambdab{\lambda\kern-1.3mm\hbox{\vrule width1mm height 1.5mm depth -1.4mm}}
\def\slambdab{\lambda\kern-1.16mm\hbox{\vrule width0.8mm height 1.1mm depth -1.mm}}
\def\sslambdab{\lambda\kern-1.04mm\hbox{\vrule width0.6mm height 0.82mm depth -0.7mm}}



\mathchardef\lap='0001


\def\lsim{\setbox0=\hbox{$\displaystyle 
\raise2.2pt\hbox{$\;<$}\kern-7.7pt\lower2.6pt\hbox{$\sim$}\;$}
\box0}
\def\gsim{\setbox0=\hbox{$\displaystyle 
\raise2.2pt\hbox{$\;>$}\kern-7.7pt\lower2.6pt\hbox{$\sim$}\;$}
\box0}



\def\gammae{\gamma_{\rm E}}













\def\frac#1#2{{#1\over#2}}
\def\dfrac#1#2{{\displaystyle{#1\over#2}}}
\def\tfrac#1#2{{\textstyle{#1\over#2}}}
\def\ffrac#1#2{\leavevmode
   \kern.1em \raise .5ex \hbox{\the\scriptfont0 #1}%
   \kern-.1em $/$%
   \kern-.15em \lower .25ex \hbox{\the\scriptfont0 #2}%
}%

\def\oversetbrace#1\to#2{\overbrace{#2}^{#1}}
\def\undersetbrace#1\to#2{\underbrace{#2}_{#1}}



\def\rightcorner#1#2{\vrule height-2.4pt width#2 depth2.9pt
\vrule height #1 depth2.8pt width.5pt\kern-.6mm}

\def\rightroof#1#2{\vrule height 9pt depth -8.5pt width#2
\vrule height 8.8pt depth#1 width.5pt
\kern-.6mm}


\def\lcorner{\kern0.8mm\vrule height 1mm \phantom{.}\kern-1mm\vrule height0.4pt width1mm}
\def\rcorner{\vrule height0.4pt width0.1mm\kern-1mm\phantom{.}\vrule height 1mm}

\def\tlcorner{\kern0.8mm\vrule height 0.53mm \phantom{.}\kern-1mm\vrule height0.4pt width1mm}
\def\trcorner{\vrule height0.4pt width0.1mm\kern-1mm\phantom{.}\vrule height 0.53mm}

\def\upcorchfill{
$\lcorner\leaders\vrule height0.4pt\hfill
\leaders\vrule height0.4pt\hfill\rcorner\kern0.8mm $}

\def\tupcorchfill{
$\tlcorner\leaders\vrule height0.4pt\hfill
\leaders\vrule height0.4pt\hfill\trcorner\kern0.8mm $}

\def\dundercontraction#1{\mathop{\vtop{\ialign{##\crcr
$\hfil\kern-.2mm{\displaystyle #1}\hfil$\crcr
\noalign{\kern2pt\nointerlineskip}\upcorchfill\crcr\noalign{\kern3pt}}}}\!\,}

\def\tundercontraction#1{\mathop{\vtop{\ialign{##\crcr
$\hfil\kern-.2mm{\displaystyle #1}\hfil$\crcr
\noalign{\kern2pt\nointerlineskip}\tupcorchfill\crcr\noalign{\kern3pt}}}}\!\,}

\def\sundercontraction#1{\mathop{\vtop{\ialign{##\crcr
$\hfil\kern-.2mm{\scriptstyle #1}\hfil$\crcr
\noalign{\kern2pt\nointerlineskip}\tupcorchfill\crcr\noalign{\kern3pt}}}}\!\,}
                
\def\ssundercontraction#1{\mathop{\vtop{\ialign{##\crcr
$\hfil\kern-.2mm{\scriptscriptstyle #1}\hfil$\crcr
\noalign{\kern2pt\nointerlineskip}\tupcorchfill\crcr\noalign{\kern3pt}}}}\!\,}


\def\dundercontractionlimits#1{\mathop{\vtop{\ialign{##\crcr
$\hfil\kern-.2mm\displaystyle{#1}\hfil$\crcr
\noalign{\kern3pt\nointerlineskip}
\upcorchfill\crcr\noalign{\kern3pt}}}}\limits}


\def\yundercorch#1#2#3{\mathord{\vtop to0pt{\ialign{##\crcr
$\hfil\displaystyle{#3}\hfil$\crcr\noalign{\kern1.5pt\nointerlineskip}
$\hfil\ycorch{#1}{#2}{}\kern1.5pt\hfil$\crcr}\vss}}}

\def\ycorch#1#2{\vrule height #1 \vrule height 0.4pt width #2\vrule height #1}

\def\youundercontract#1#2{\;\yundercorch{#1}{#2}{\phantom{,}}\kern-4pt\kern-#2}


\def\tlroof{\vrule height0.1mm depth 0.55mm \hbox{\vrule height0.4pt width 1mm}}
\def\trroof{\hbox{\vrule height0.4pt width1.mm}\vrule height 0.1mm depth0.55mm}

\def\lroof{\vrule height0.1mm depth 1mm \hbox{\vrule height0.4pt width 1mm}}
\def\rroof{\hbox{\vrule height0.4pt width1.mm}\vrule height 0.1mm depth1mm}

\def\tovercorchfill{
$\tlroof\leaders\vrule height0.4pt depth0mm\hfill
\leaders\vrule height0.4pt depth 0mm\hfill\trroof\kern0.8mm $}

\def\overcorchfill{
$\lroof\leaders\vrule height0.4pt depth0mm\hfill
\leaders\vrule height0.4pt depth 0mm\hfill\rroof\kern0.8mm $}

\def\dovercontraction#1{\mathop{\vbox{\ialign{##\crcr\noalign{\kern3pt}
\kern.6mm\overcorchfill\crcr\noalign{\kern2pt\nointerlineskip}
$\hfil\kern-0.4mm{\displaystyle #1}\hfil$\crcr}}}\!\,}

\def\tovercontraction#1{\mathop{\vbox{\ialign{##\crcr\noalign{\kern3pt}
\kern.6mm\tovercorchfill\crcr\noalign{\kern2pt\nointerlineskip}
$\hfil\kern-0.4mm{\displaystyle #1}\hfil$\crcr}}}\!\,}

\def\sovercontraction#1{\mathop{\vbox{\ialign{##\crcr\noalign{\kern3pt}
\kern.6mm\tovercorchfill\crcr\noalign{\kern2pt\nointerlineskip}
$\hfil\kern-0.4mm{\scriptstyle #1}\hfil$\crcr}}}\!\,}

\def\ssovercontraction#1{\mathop{\vbox{\ialign{##\crcr\noalign{\kern3pt}
\kern.6mm\tovercorchfill\crcr\noalign{\kern2pt\nointerlineskip}
$\hfil\kern-0.4mm{\scriptscriptstyle #1}\hfil$\crcr}}}\!\,}


\def\dovercontractionlimits#1{\mathop{\vbox{\ialign{##\crcr\noalign{\kern3pt}
\overcorchfill\crcr\noalign{\kern3pt\nointerlineskip}
$\hfil\displaystyle{#1}\hfil$\crcr}}}\limits}


\def\yovercorch#1#2{\kern2pt\vrule height 0.1mm depth #1 \hbox{\vrule height 0.4pt width #2}
\kern-1.1mm\vrule height0.1mm depth #1}

\def\yabovecorch#1#2#3{\mathop{\vbox{\ialign{##\crcr\noalign{\kern3pt}
\yovercorch{#1}{#2}\crcr\noalign{\kern3pt\nointerlineskip}
$\hfil\displaystyle{#3}\hfil$\crcr}}}\limits}

\def\youovercontract#1#2{\yabovecorch{#1}{#2}{\phantom{I}}\kern-6pt\kern-#2}



\def\yoverc#1#2#3{\kern2pt\vrule height 2.6pt depth #1
 \hbox{\vrule height 2.6pt depth-2.1pt width #2}
\kern-4.2mm$\rightarrow#3$}
\def\yabovec#1#2#3#4{\mathop{\vbox{\ialign{##\crcr\noalign{\kern3pt}
\yoverc{#1}{#2}{#3}\crcr\noalign{\kern3pt\nointerlineskip}
$\displaystyle{#4}\hfil$\crcr}}}\limits}


\def\yc#1#2#3{\vrule height #1depth-2.4pt \vrule height 2.75pt depth-2.25pt width #2\kern-2.9mm \rightarrow #3}
\def\yunderc#1#2#3#4{\mathord{\vtop to0pt{\ialign{##\crcr
$\hfil\displaystyle{#4}\hfil$\crcr\noalign{\kern1.5pt\nointerlineskip}
$\hfil\yc{#1}{#2}{#3}\hfil$\crcr}\vss}}}

\def\underdeviatea#1#2#3#4{\setbox0=\hbox{$#3$}\;\,\yunderc{#1}{#2}{\box0}{\kern -\wd0 #4}\nullbox{0mm}{#1}{0mm}}


\def\yovercna#1#2#3{\kern2pt\vrule height 2.6pt depth #1
 \hbox{\vrule height 2.6pt depth-2.1pt width #2}
$#3$}
\def\yabovecna#1#2#3#4{\mathop{\vbox{\ialign{##\crcr\noalign{\kern3pt}
\yovercna{#1}{#2}{#3}\crcr\noalign{\kern3pt\nointerlineskip}
$\displaystyle{#4}\hfil$\crcr}}}\limits}


\def\ycna#1#2#3{\vrule height #1depth-2.4pt
 \vrule height 2.75pt depth-2.25pt width #2 #3}
\def\yundercna#1#2#3#4{\mathord{\vtop to0pt{\ialign{##\crcr
$\hfil\displaystyle{#4}\hfil$\crcr\noalign{\kern0pt\nointerlineskip}
$\hfil\kern-1.2em\ycna{#1}{#2}{#3}\hfil$\crcr}\vss}}}

\def\underdeviate#1#2#3#4{
\setbox0=\hbox{$\;#3$}\kern1.4em\yundercna{#1}{#2}{\box0}{\kern -\wd0 #4}\nullbox{0mm}{#1}{0mm}}






\font\smallsc=cmcsc10 at 9pt 
\font\fib=cmfib8
\font\medfib=cmfib8 at 9pt
\font\bigfib=cmfib8 at 12pt




\font\addressfont=cmbxti10 at 9pt




\def\brochureb#1#2#3{\pageno#3
\headline={\ifodd\pageno{\ifnum\pageno=#3\hfil\else\rheadline\fi}
\else\lheadline\fi}
\def\rheadline{\hfil -{#2}-\hfil}
\def\lheadline{\hfil-{#1}-\hfil}
\footline={\ifnum\pageno=#3\hfil
\else\advance\pageno by-1\hss -- \number\pageno\ --\hss\fi}
\voffset=2\baselineskip}


\def\brochuret#1#2#3{\pageno#3
\headline={\ifodd\pageno{\ifnum\pageno=#3\hfil}
\else\advance\pageno by -1\hss\trheadline\fi}
\def\trheadline{\hfil -{#2}-\hfil\folio}
\def\tlheadline{\folio\hfil-{#1}-\hfil}
\nopagenumbers
\voffset=2\baselineskip}
\def\nada{\phantom{M}\kern-1em}
\def\brochureendcover#1{\vfill\eject\pageno=1{\nada#1}\vfill\eject}



\def\brochureendchapter{\ifodd\pageno\vfill\eject
 \else\vfill\eject\null \vfill\eject \fi}



\def\chapterb#1#2#3{\pageno#3
\headline={\ifodd\pageno{\ifnum\pageno=#3\hfil\else\rheadline\fi}
\else\lheadline\fi}
\def\rheadline{\hfil -{#2}-\hfil}
\def\lheadline{\hfil-{#1}-\hfil}
\footline={\hss -- \number\pageno\ --\hss}
\voffset=2\baselineskip}


\def\chaptert#1#2#3{\pageno#3
\headline={\ifodd\pageno{\ifnum\pageno=#3\hfil\folio\else\trheadline\fi}
\else\tlheadline\fi}
\def\trheadline{\hfil -{#2}-\hfil\folio}
\def\tlheadline{\folio\hfil-{#1}-\hfil}
\nopagenumbers
\voffset=2\baselineskip}

\def\bookendchapter{\ifodd\pageno\vfill\eject\null \vfill\eject
 \else\vfill\eject \fi}



\def\brochuresection#1{\setbox0=\vbox{
\hsize=0.8\hsize\raggedright\tolerance=400\hfuzz=4mm
\medbreak\noindent{\medfib #1}}\smallskip\box0
\nobreak
\noindent}








\def\abstracttype#1{\hsize0.7\hsize\tolerance=800\hfuzz=0.5mm \noindent{\fib #1}\par
\medskip\petit}


\def\hb{\hfill\break}







\magnification1000
\brochureb{\smallsc a. pineda and f. j.  yndur\'ain}{\smallsc comment on 
``calculation of quarkonium 
spectrum and $m_b,\,m_c$ to order $\alpha_s^4$''}{1}
\rightline{FTUAM 98-27}
\rightline{CERN-TH/98-402}
\rightline{December, 14, 1998}
\bigskip
\hrule height .3mm
\vskip.6cm
\centerline{{\bigfib Comment on ``Calculation of Quarkonium Spectrum and $m_b,\,m_c$
 to Order $\alpha_s^4$''}\footnote*{\petit Supported in part by CICYT, Spain}}
\medskip
\centerrule{.7cm}
\vskip1cm
\setbox8=\vbox{\hsize65mm {\noindent\fib A. Pineda}
\vskip .1cm
\noindent{\addressfont Theory Division\hb
CERN,\hb
1211 Geneva 23, Switzerland}}
\centerline{\box8}
\smallskip
\setbox7=\vbox{\hsize65mm \fib and} 
\centerline{\box7}
\smallskip
\setbox9=\vbox{\hsize65mm {\noindent\fib F. J. 
Yndur\'ain} 
\vskip .1cm
\noindent{\addressfont Departamento de F\'{\i}sica Te\'orica, C-XI,\hb
 Universidad Aut\'onoma de Madrid,\hb
 Canto Blanco,\hb
E-28049, Madrid, Spain.}\hb}
\smallskip
\centerline{\box9}
\bigskip
\setbox0=\vbox{\abstracttype{Abstract}In a recent paper, we included
 two loop, relativistic one loop and second order
 relativistic tree level corrections, plus leading 
 nonperturbative contributions, to obtain a calculation of the lower states in the heavy
 quarkonium spectrum correct 
up to, and including, $O(\alpha_s^4)$ and leading $\Lambdav^4/m^4$ terms.
The results were obtained with, in particular, the value of the two loop static coefficient
 due to Peter; this been recently challenged by Schr\"oder. In our 
previous paper we used Peter's result; in the present one we now give 
results with Schr\"oder's, as this is likely to be 
the correct one. The 
variation is  slight as the value of $b_1$ is only 
one among the various $O(\alpha_s^4)$ contributions.
 With Schr\"oder's expression we now have,
$$m_b=5\,001^{+104}_{-66}\;\mev;\quad
\bar{m}_b(\bar{m}_b^2)=4\,440^{+43}_{-28}\;\mev,$$
$$m_c=1\,866^{+190}_{-154}\;\mev;\quad
\bar{m}_c(\bar{m}_c^2)=1\,531^{+132}_{-127}\;\mev.$$
Moreover,
$$\Gammav(\Upsilonv\rightarrow e^+e^-)=1.07\pm0.28\;\kev
\;(\hbox{exp.}=1.320\pm0.04\,\kev)$$
and the hyperfine splitting is predicted to be
$$M(\Upsilonv)-M(\eta)=47^{+15}_{-13}\;\mev.$$}
\centerline{\box0}
\brochureendcover{Typeset with \physmatex}

\brochuresection{1 Introduction}
In a recent paper\ref{1}, to be hereafter denoted by PYI, we 
evaluated the heavy quarkonium spectrum to order $\alpha_s^4$. 
The ingredients for the calculation were the one loop corrections\ref{2} to the static potential, 
and its first and second order contributions to the spectrum; the relativistic, 
and mixed relativistic--one loop\ref{3} corrections;  
the two loop corrections to the static potential, and the leading 
 nonperturbative corrections\ref{4}. The 
two loop correction to the static potential used was that
 calculated by Peter\ref{5}; but the calculation of Peter 
has   recently been 
challenged by Schr\"oder\ref{6}. In this last 
reference, the result of Peter's is 
checked for all pieces except one of the contributions
 to the $C_A^2$ coefficient, where an error in Peter's evaluation is pointed out. 
In the present paper we give the 
results of the calculation 
using the Schr\"oder results. Note that the variation is {\sl very} small; 
the reason is that the two loop static potential is only one of the several contributions to the 
$O(\alpha_s^2)$ calculation. For example, 
Using the value of the two loop 
coefficient found by Peter, we found
$$m_b=5\,015^{+110}_{-70}\;\mev;\;
m_c=1\,884^{+222}_{-133}\;\mev\quad(P)\equn{(1.1a)}$$
to which correspond the $\overline{\hbox{MS}}$ masses,
$$\bar{m}_b(\bar{m}_b^2)=4\,453^{+50}_{-32}\;\mev;\;
\bar{m}_c(\bar{m}_c^2)=1\,547^{+169}_{-102}\;\mev.\quad(P)$$

With Schr\"oder's result, one now has
$$\eqalign{m_b=5\,001^{+104}_{-66}\;\mev;\phantom{\quad(S)}\cr
\bar{m}_b(\bar{m}_b^2)=4\,440^{+43}_{-28}\;\mev\quad(S)\cr}\equn{(1.1b)}$$
and, for the $c$ quark,
$$\eqalign{m_c=1\,866^{+190}_{-154}\;\mev;\phantom{\quad(S)}\cr
\bar{m}_c(\bar{m}_c^2)=1\,531^{+132}_{-127}\;\mev\quad(S)\cr}\equn{(1.1c)}$$

For the leptonic decay of the $\Upsilonv$, and the hyperfine splitting 
the results with Schr\"oder,s value of $b_1$ are
$$\Gammav(\Upsilonv\rightarrow e^+e^-)=1.07\pm0.28\;\kev
\;(\hbox{exp.}=1.320\pm0.04\,\kev)$$
and
$$M(\Upsilonv)-M(\eta)=46.6^{+14.8}_{-12.7}\;\mev$$
they are almost identical to those obtained before.

\brochuresection{2 The Effective Potential}
We follow the method of effective potentials, and the
 renormalization scheme of ref.~3. The Hamiltonian for quarkonium is 
$$H=H^{(0)}+H_1\equn{(2.1a)}$$
where 
$$\eqalign{H^{(0)}=&2m+\dfrac{-1}{m}\lap-\dfrac{C_F\tilde{\alpha}_s(\mu^2)}{r},\cr
\tilde{\alpha}_s(\mu^2)=&\alpha_s(\mu^2)
\left\{1+\left(a_1+\dfrac{\gammae\beta_0}{2}\right)\dfrac{\alpha_s(\mu^2)}{\pi}\right.\cr
&\left.+\left[\gammae\left(a_1\beta_0+\dfrac{\beta_1}{8}\right)+
\left(\dfrac{\pi^2}{12}+\gammae^2\right)\dfrac{\beta_0^2}{4}+
b_1\right]\dfrac{\alpha_s^2}{\pi^2}\right\}\cr}\equn{(2.1b)}$$
and  will be solved exactly. $H_1$ is
$$H_1=V_{\rm tree}+V^{(L)}_1+V^{(L)}_2+V^{(LL)}+V_{\rm s.rel}+V_{\rm spin},\equn{(2.1c)}$$
and (we repeat the formulas of PYI for ease of reference)
$$\eqalign{V_{\rm tree}=&\dfrac{-1}{4m^3}\lap^2+\dfrac{C_F\alpha_s}{m^2r}\lap,\cr
V^{(L)}_1=&\dfrac{-C_F\beta_0\alpha_s(\mu^2)^2}{2\pi}\,\dfrac{\log r\mu}{r},\cr
V^{(L)}_2=&\dfrac{-C_F\alpha_s^3}{\pi^2}\,
\left(a_1\beta_0+\dfrac{\beta_1}{8}+\dfrac{\gammae\beta_0^2}{2}\right)\dfrac{\log r\mu}{r}\cr
&\equiv\dfrac{-C_Fc_2^{(L)}\alpha_s^3}{\pi^2}\;\dfrac{\log r\mu}{r},\cr
V^{(LL)}=&\dfrac{-C_F\beta_0^2\alpha_s^3}{4\pi^2}\;\dfrac{\log^2 r\mu}{r},\cr
V_{\rm s.rel}=&\dfrac{C_Fa_2\alpha_s^2}{2mr^2},\cr
V_{\rm spin}=&\dfrac{4\pi C_F\alpha_s}{3m^2}s(s+1)\delta({\bf r}).\cr}\equn{(2.1d)}$$
Here the running coupling constant has to be taken to three loops. 
 $a_1$ was calculated in 
ref.~1 and $a_2$ in ref.~3. Both have been checked 
by independent 
calculations; the only discrepancy lies in the value of 
the coefficient \fnote{For
 the values of the constants other than $b_1$ 
entering above formulas, cf. PYI.} $b_1$. According to Peter,\ref{5}
$$\eqalign{b_1=&\tfrac{1}{16}
\Big\{\left[\tfrac{4343}{162}+6\pi^2-\tfrac{1}{4}\pi^4+\tfrac{22}{3}\zeta(3)\right]C_A^2\cr
&-\left[\tfrac{1798}{81}+\tfrac{56}{3}\zeta(3)\right]C_AT_Fn_f-
\left[\tfrac{55}{3}-16\zeta(3)\right]C_FT_Fn_f+\tfrac{400}{81}T_F^2n_f^2\Big\}\cr
&\simeq 24.30, \cr}\equn{(2.2a)} $$
while Schr\"oder\ref{6} gives
$$\eqalign{b_1=&\tfrac{1}{16}
\Big\{\left[\tfrac{4343}{162}+4\pi^2-\tfrac{1}{4}\pi^4+\tfrac{22}{3}\zeta(3)\right]C_A^2\cr
&-\left[\tfrac{1798}{81}+\tfrac{56}{3}\zeta(3)\right]C_AT_Fn_f-
\left[\tfrac{55}{3}-16\zeta(3)\right]C_FT_Fn_f+\tfrac{400}{81}T_F^2n_f^2\Big\}\cr
&\simeq 13.2. \cr}\equn{(2.2b)} $$

Although the difference between the two 
lies only in one piece of the 
coefficient of $\tfrac{1}{16}C_F^2$, a $4\pi^2$ vs. a $6\pi^2$, 
the Schr\"oder result makes the two loop correction much smaller (and hence the 
overall calculation more believable). We present here results with (2.2b) 
since they are the ones more likely to be correct (moreover, results with (2.2a) 
can be found in PYI).

It should be noted that $H_1$ contains a tree level velocity correction, and a 
velocity-dependent one
loop piece, $V_{\rm s.rel}$. This is because the average velocity in a Coulombic potential 
is $\langle |v|\rangle\sim \alpha_s$, hence 
a calculation correct to order $\alpha_s^4$ requires tree level $O(v^2)$ and 
one loop $O(|v|)$ contributions. 
All these terms in $H_1$ may be treated as perturbations to first order, {\sl except} 
$V^{(L)}_1$. 
For this, the second order perturbative contribution is also required as this
 also produces a correction of order $\alpha_s^4$. It is  to be noted that {\sl all} 
the dependence on $b_1$ is contained in the $b_1$ dependence of $\widetilde{\alpha}_s$, 
\equ~(2.1b).

A last comment concerns the renormalization scheme. We have followed ref.~3 
in renormalizing $\alpha_s$ in the \msbar\ scheme; but the mass $m$ that
 appears in \equs~(2.1) is the two loop pole mass. That is to say, it is defined by the 
equation,
$$S^{-1}_2(\slash{p}=m,m)=0\equn{(2.3)}$$
where $S_2(\slash{p},m)$ is the quark propagator to two loops. One can relate $m$ 
to the \msbar\ parameter, also to two loop accuracy, using the results of refs.~7.

Nonperturbative corrections are not included 
in \equs{(2.1)}; they will be incorporated  later.

\brochuresection{3 Energy Shifts, Order $\alpha_s^4,\,\Lambdav^4/m^4$}
Taking into account the expression for the Hamiltonian, \equn{(2.1)}, we  write 
$$E_{nl}=2m-m\dfrac{C_F^2\tilde{\alpha}_s^2}{4n^2}+\sum_{V}\delta^{(1)}_{V}E_{nl}
+\delta^{(2)}_{V_1^{(L)}}E_{nl}+\delta_{\rm NP}E_{nl}.\equn{(3.1)}$$
We define generally the analogue of the Bohr radius,
$$a(\mu^2)=\dfrac{2}{mC_F\tilde{\alpha}_s(\mu^2)},$$
and then (PYI),
$$\delta^{(1)}_{V_{\rm tree}}E_{nl}=-\dfrac{2}{n^3\,m^3\,a^4}
\left[\dfrac{1}{2l+1}-\dfrac{3}{8n}\right]+
\dfrac{C_F\alpha_s}{m^2}\,\dfrac{2l+1-4n}{n^4(2l+1)a^3};\equn{(3.2a)}$$
$$\delta^{(1)}_{V^{(L)}_1}E_{nl}=
-\dfrac{\beta_0C_F\alpha^2_s(\mu^2)}{2\pi n^2a}
\left[\log\dfrac{na\mu}{2}+\psi(n+l+1)\right];\equn{(3.2b)}$$
$$\delta^{(1)}_{V_2^{(L)}}E_{nl}=-\dfrac{C_Fc_2^{(L)}\alpha_s^3}{\pi^2n^2a}\;
\left[\log\dfrac{na\mu}{2}+\psi(n+l+1)\right];\equn{(3.2c)}$$
$$\eqalign{\delta^{(1)}_{V^{(LL)}}E_{nl}=-\dfrac{C_F\beta_0^2\alpha_s^3}{4\pi^2n^2a}\,
\Big\{\log^2\dfrac{na\mu}{2}+2\psi(n+l+1)\log\dfrac{na\mu}{2}\cr
+\psi(n+l+1)^2+\psi'(n+l+1)\cr
+\theta(n-l-2)\dfrac{2\Gammav(n-l)}{\Gammav(n+l+1)}
\sum^{n-l-2}_{j=0}\dfrac{\Gammav(2l+2+j)}{j!(n-l-j-1)^2}\Big\};\cr}
\equn{(3.2d)}$$
$$\delta^{(1)}_{V_{\rm s.rel}}E_{nl}=\dfrac{C_Fa_2\alpha_s^2}{m}\;
\dfrac{1}{n^3(2l+1)a^2}.\equn{(3.2e)}$$
 For the vector states ($\Upsilonv,\,\Upsilonv',\,\Upsilonv'';\;J/\psi,\,\psi',\dots$) 
one has to add the hyperfine shift, at tree level,
$$\delta^{(1)}_{V_{\rm spin}}E_{nl}=\delta_{s1}\delta_{l0}
\dfrac{8C_F\alpha_s}{3n^3m^2a^3}.\equn{(3.2f)}$$
The calculation of the second order 
contribution of $V_1^{(L)}$,  $\delta^{(2)}_{V_1^{(L)}}E_{nl}$, is nontrivial,
 and may be found in PYI. We write
$$\delta^{(2)}_{V_1^{(L)}}E_{nl}\equiv 
-m\dfrac{C_F^2\beta_0^2\alpha_s^4}{4n^2\pi^2}
\left\{N_0^{(n,l)}+N_1^{(n,l)}\log \dfrac{na\mu}{2}+
\tfrac{1}{4}\log^2\dfrac{na\mu}{2}\right\} \equn{(3.3a)}$$
and the $N$ are given in PYI. These equations are unchanged from PYI;
 we repeat them here for ease of reference. 
 
The dominant nonperturbative 
corrections  are associated with the gluon condensate and are\ref{4}
$$\eqalign{\delta_{\rm NP}E_{nl}=
m\epsilon_{nl}n^2\pi\langle\alpha_sG^2\rangle\left(\dfrac{na}{2}\right)^4=
m\dfrac{\epsilon_{nl}n^6\pi\langle\alpha_sG^2\rangle}{(mC_F\tilde{\alpha}_s)^4};\cr
\epsilon_{10}=\tfrac{1\,872}{1\,275},\;\epsilon_{20}=\tfrac{2\,102}{1\,326},\;
\epsilon_{21}=\tfrac{9\,929}{9\,945}.\cr}\equn{(3.4)}$$

Because $\langle \alpha_sG^2\rangle\sim\Lambdav^4$, this is of order 
$(\Lambdav/m)^4$. Besides the corrections reported above, there are 
a few pieces of the higher order perturbative and nonperturbative 
corrections that are known; 
they can be found discussed in PYI. 

\brochuresection{4 Numerical Results}
Using the formulas deduced above one evaluates the quark mass,  and spectrum 
and other properties of heavy 
quarkonium systems. 
We take,
$$\Lambdav(n_f=4,\,\hbox{three loops})=0.23^{+0.08}_{-0.05}\;\gev\;
\left[\;\alpha_s(M_Z^2)\simeq0.114^{+0.006}_{-0.004}\;\right],
\equn{(4.1a)}$$ 
and for the gluon condensate, very poorly known,
$$\langle\alpha_sG^2\rangle=0.06\pm0.02\;\gev^4.\equn{(4.1b)}$$

This value of $\alpha_s(M_Z^2)$ is slightly smaller 
than, though compatible with, the world average $\alpha_s(M_Z^2)=0.118$. We have 
preferred our value, which is obtained averaging measurements 
performed at {\sl spacelike} momenta; see the 
recent review of S. Bethke\ref{8}.

Another matter to be discussed is the choice of the renormalization 
point, $\mu$. As our equations (3.2, 3) show, a {\sl natural} value for this 
parameter is 
$$\mu=\dfrac{2}{na},\equn{(4.2)}$$
for states with the principal quantum number $n$, and this will be our choice. For 
states with $n=1$ the results of the 
calculation will turn out to depend  little on the value of 
$\mu$, provided it is reasonably close to (4.2). Higher states are 
another matter; we will discuss our choices when we consider them.
\medbreak
\noindent{\fib The 10 state of $\bar{b}b$ and the mass $m_b$}.\quad 
As stated, we select, for the $\Upsilonv$ state, $\mu=2/a$. We then use \equs~(3.1-4) 
to obtain the values of the $b$ quark mass.  The results are reported  below;
the errors correspond to the errors in \equs~(4.1a, b). 
In the estimate of the errors, the condition $\mu=2/a$ is 
maintained satisfied when varying $\Lambdav$ while for the error 
due to the variation of $\mu$ the other parameters are kept fixed (i.e., one no more 
has then $\mu=2/a$). The dependence of $m_b$ on $\mu$ should 
be taken as an indication of the theoretical 
uncertainty of our calculation. With  Schr\"oder's value for $b_1$,  
$$\eqalign{m_b=5.001^{+0.097}_{-0.061}\,(\Lambdav)\;\mp0.005\,(\langle\alpha_sG^2\rangle)\;
^{-0.025}_{+0.037} \;(\hbox{vary}\; \mu^2\;{\rm by}\,25\%)
\;\pm 0.006\;({\rm other\; th.\;uncertainty})\cr
\bar{m}_b(\bar{m}_b^2)=4.440^{+0.025}_{-0.013}\,(\Lambdav)
\,\mp0.005\,(\langle\alpha_sG^2\rangle)\;
^{-0.023}_{+0.035} \;(\hbox{vary}\; \mu^2\;{\rm by}\,25\%)
\;\pm0.005\;({\rm other\; th.\;uncertainty}).\cr}
\equn{(4.3)}$$
The values of $\mu^2$, $\alpha_s(\mu^2)$, $\tilde{\alpha}_s(\mu^2)$ are, respectively, 
$$\mu^2=6.632\,\gev^2,\; \alpha_s(\mu^2)=0.246\;, \tilde{\alpha}_s(\mu^2)=0.386.$$

The piece denoted by the expression ``other th. uncertainty" in (4.3) 
refers to the error coming from 
higher dimensional operators and higher order perturbative terms; 
it can be found discussed in PYI. It 
 is comfortably 
smaller that the errors due to the uncertainty on $\Lambdav,\;\langle\alpha_sG^2\rangle$.
If we  omit these errors, so as not to double count them,
 and consider that the theoretical 
error is only that due to varying $\mu^2$ by 25\%, and 
 compose all the errors quadratically, then we obtain the estimates reported in 
the Introduction, \equs~(1.1).
\medbreak
\noindent{\fib $M(\Upsilonv)-M(\eta_b)$; the decay
 $\Upsilonv\rightarrow e^+e^-$}.\quad  The expressions 
for the hyperfine splitting, and the decay of
 the $\Upsilonv$ into $e^+e^-$ are as in PYI. They depend on $b_1$ 
only indirectly, through the 
preferrred values of $m$, $\mu$. 
We have the numerical results, using the Schr\"oder calculation
$$M(\Upsilonv)-M(\eta)=46.6^{+10.9}_{-3.5}\,(\Lambdav)\,^{+5.5}_{-5.2}\,
(\langle\alpha_sG^2\rangle)\,
^{+8.3}_{-11.1}\,(\mu^2=6.632\pm25\%)\equn{(4.4)}$$
and
$$\Gammav(\Upsilonv\rightarrow e^+e^-)=1.07^{+0.11}_{+0.01}\,(\Lambdav)\,
^{+0.12}_{-0.11}\,(\langle\alpha_s G^2\rangle)\,^{+0.21}_{-0.26}\,(\mu^2=6.632\pm25\%).\equn{(4.5)}$$
practically unchanged from PYI.
Note that, when varying $\Lambdav,\;\langle \alpha_sG^2\rangle$, 
we have varied $m_b$ according to \equn{(4.3)}, but we have {\sl not} 
varied $m_b$ when varying $\mu$. Note also that the corrections are here fairly large; 
in particular, due to the large size of the radiative correction to the decay.\ref{9}

 Higher order NP corrections due to the higher dimensional operators 
are also known for the decay rate (see PYI). 
Size corrections, however, are not known now.

The result for the decay is in reasonable agreement with experiment,
$$\Gammav_{\rm exp.}(\Upsilonv\rightarrow e^+e^-)=1.320\pm0.04\,\kev.$$
\medbreak
\noindent{\fib  Higher states ($n=2$) of $\bar{b}b$}.\quad 
 As is 
clear from the expressions (3.2, 3) the natural choice of scale is now 
$\mu=1/a=2.860\;\gev^2$.  If we take this, adding or subtracting 
a 25\% to estimate the dependence of the calculation on the choice of scale then we obtain  
 the results, with  the Schr\"oder value of $b_1$ 
are,
$$
\eqalign{M(20,\;\hbox{th})-M(20,\;\hbox{exp})=363^{+310}_{-324}\;\mev\;
(\mu^2=2.86\pm25\%),\cr
M(21,\;\hbox{th})-M(21,\;\hbox{exp})=208^{+205}_{-216}\;\mev\;(\mu^2=2.86
\pm25\%).\cr}\equn{(4.6)}
$$
\medbreak 
\noindent{\fib The 10 state of $\bar{c}c$ and the mass $m_c$}.\quad
The value of the parameter $\Lambdav$ used now, corresponding to that in \equn{(4.1a)}, is
$$\Lambdav(n_f=3,\,\hbox{three loops})=0.30^{+0.09}_{-0.05}\;\gev.$$
The values for the $c$ quark mass, deduced from the $J/\psi$ mass 
are now,
$$\eqalign{m_c=1.866^{+0.154}_{-0.091}\,(\Lambdav)\;\mp0.014\,(\langle\alpha _sG^2\rangle)\;
^{-0.124}_{+0.110} \;(\hbox{varying}\; \mu^2\;{\rm by}\,25\%)\;\pm\;0.011\;
({\rm th.\;uncertainty})\phantom{\quad(P)}\cr
\bar{m}_c(\bar{m}_c^2)=
1.531^{+0.083}_{-0.052 }\,(\Lambdav)\,\mp 0.013\,(\langle\alpha_sG^2\rangle)\;
^{-0.115}_{+0.102} \;(\hbox{varying}\; \mu^2\;{\rm by}\,25\%)\;\pm\;0.010\;
({\rm th.\;uncertainty})\quad(S),\cr}
\equn{(4.7)}$$
and $\mu^2=2.465\;\gev^2$ now. 
Composing the errors as for the $b$-quark we find the results 
reported in the Introduction, \equs~(1.1). 

\brochuresection{5 Discussion}
The discussion of PYI holds valid for 
the calculations using both Peter's and Schr\"oder's evaluations, with 
one point of difference. If we believe Schr\"oder's value of $b_1$, 
then the two loop corrections are comfortably small. For example, with Peter's value we had
$$a_1\alpha_s/\pi\simeq0.11,\quad b_1\alpha_s^2/\pi^2\simeq0.14,\quad(P)$$
while with Schr\"oder's the first is almost unchanged but the second becomes 
$\quad b_1\alpha_s^2/\pi^2\simeq0.081\quad(S)$
\bigskip

\brochuresection{Acknowledgments}
A.P. acknowledges the support of the European Community, Marie Curie fellowship, TMR Contract n. ERBFMBICT983405.

\brochuresection{References}
\item{1}{A. Pineda and F. J. Yndur\'ain, Phys. Rev. {\bf D58} (1998) 094022.
\item{2}{W. Fischler, Nucl. Phys. {\bf B129} (1977) 157; A. Billoire, Phys. Lett. {\bf B92} (1980) 343.}
\item{3}{S. Titard and F. J. Yndur\'ain, Phys. Rev. {\bf D49} (1994) 6007.}
\item{4}{M. B. Voloshin, Nucl. Phys. {\bf B154} (1979) 365 and Sov. J. Nucl. Phys. 
{\bf 36} (1982) 143; H. Leutwyler, Phys. Lett. {\bf B98} (1981) 447;
Yu. A. Simonov, S. Titard and  F. J. Yndur\'ain, Phys. Lett. {\bf B354} (1995) 435; 
S. Titard and F. J. Yndur\'ain, Phys. Rev. {\bf D51} (1995) 6348;
the analysis corrected for some states  
by A. Pineda,  Phys. Rev. {\bf D55} (1997) 407; 
A. Pineda, Nucl. Phys. {\bf B494} (1997) 213.}
\item{5}{M. Peter, Phys. Rev. Lett. {\bf 78} (1997) 602.}
\item{6}{Y. Schr\"oder, DESY 98-191 (hep-ph/9812205), 1998.}
\item{7}{R. Coquereaux, Phys. Rev. {\bf D23} (1981) 1365 to one loop;
N. Gray et al., Z. Phys. {\bf C48} (1990) 673 to two loops.}
\item{8}{S. Bethke, Nucl. Phys. Proc. Suppl. {\bf 64} (1998) 54.} 
\item{9}{R. Barbieri et al., Phys. Lett. {\bf 57B} (1975) 455; {\sl ibid.} Nucl. 
Phys. {\bf B154} (1979) 535.}

\bye